\documentclass[preprint,showpacs,showkeys,preprintnumbers,amsmath,amssymb]{revtex4}
\usepackage{dcolumn}
\usepackage{bm}

\renewcommand{\a}{\alpha}
\renewcommand{\b}{\beta}
\newcommand{\ga}{\gamma}
\newcommand{\dl}{\delta}
\newcommand{\eps}{\varepsilon}
\newcommand{\la}{\lambda}
\newcommand{\m}{\mu}
\newcommand{\n}{\nu}
\newcommand{\om}{\omega}

\renewcommand{\th}{\theta}

\newcommand{\Dl}{\Delta}

\newcommand{\Om}{\Omega}
\newcommand{\Th}{\Theta}

\newcommand{\F}{\mathcal{F}}
\newcommand{\U}{\mathcal{U}}

\newcommand{\diff}{\mathfrak{D}}

\newcommand{\id}{{\mathrm{id}}}     %% identity operation
\newcommand{\pa}{\partial}

\newcommand{\R}{\mathbb{R}}

\newcommand{\thalf}{\tfrac{1}{2}}    %% small fraction  1/2
\newcommand{\tihalf}{\tfrac{i}{2}}   %% small fraction  i/2
\newcommand{\ox}{\otimes}           %% tensor product
\newcommand{\x}{\times}
\renewcommand{\.}{\cdot}            %% another left action
\begin{document}

\preprint{}

\title{Noncommutative spacetime symmetries:\\ Twist versus covariance}

\author{J.~M. Gracia-Bond\'{\i}a}
\affiliation{Departamento de F\'{\i}sica Te\'orica I, Universidad
    Complutense de Madrid, 28040 Madrid, Spain}
\author{Fedele Lizzi}
\email{lizzi@na.infn.it}
\affiliation{Dipartimento di Scienze Fisiche, Universit\`a di Napoli
    Federico II, and \\ INFN, Sezione di Napoli, Monte S. Angelo,
    Via Cintia, 80126 Napoli, Italy}
\author{F.~Ruiz Ruiz}
\email{ferruiz@fis.ucm.es}
\affiliation{Departamento de F\'{\i}sica Te\'orica I, Universidad
    Complutense de Madrid, 28040 Madrid, Spain}
\author{Patrizia Vitale}
\email{vitale@na.infn.it}
\affiliation{Dipartimento di Scienze Fisiche, Universit\`a di Napoli
    Federico II, and \\ INFN, Sezione di Napoli, Monte S. Angelo,
    Via Cintia, 80126 Napoli, Italy}

\date{\today}

\begin{abstract}
  We prove that the Moyal product is covariant under linear affine
  spacetime transformations. From the covariance law, by introducing
  an $(x,\Theta)$-space where the spacetime coordinates and the
  noncommutativity matrix components are on the same footing, we
  obtain a noncommutative representation of the affine algebra, its
  generators being differential operators in $(x,\Theta)$-space. As a
  particular case, the Weyl Lie algebra is studied and known results
  for Weyl invariant noncommutative field theories are rederived in a
  nutshell. We also show that this covariance cannot be extended to
  spacetime transformations generated by differential operators whose
  coefficients are polynomials of order larger than one. We compare
  our approach with the twist-deformed enveloping algebra description
  of spacetime transformations.
\end{abstract}



\maketitle

\section{\label{sec:Introduction}Introduction}

It has been clear since the early days of noncommutative field theory
that Lorentz invariance, or the lack of it, plays a key role in the
subject, especially when it comes to discuss questions like causality
or unitarity. In this regard, one ought to remember there are two
distinct types of Lorentz transformations~\cite{ColladayK}. On the one
hand, there are \emph{observer} Lorentz transformations. They involve
coordinate changes under which the localized field configurations and
background fields, in this case the noncommutative parameter tensor
$\Th=(\th^{\m\n})$, transform covariantly, leaving the physics
unchanged. On the other hand, there are {\it particle} Lorentz
transformations. They involve rotations and boosts of only localized
fields within a fixed observer frame, leaving~$\Th$ unchanged and
hence modifying the physics. On the face of it, the choice of a
particular uniform background~$\Th$ breaks Lorentz symmetry to a
smaller subgroup~\cite{Iorio}, namely the group of transformations
that leave~$\Th$ unchanged. This choice is similar to the choice of a
vacuum expectation value in `spontaneous symmetry breaking'. One
should rather not speak here of unbroken/broken symmetry but rather of
hidden/manifest invariance, observer invariance accounting for hidden
and particle invariance for manifest.

The idea that noncommutative field theory is observer Poincar\'e
invariant has been around for some time~\cite{Carrolletal,Vienna}. In
ref.~\cite{Vienna} it was explicitly shown, using functional
derivative methods, that noncommutative $U(n)$ gauge theory is
observer Weyl invariant. Recall that the Weyl group includes
Poincar\'e group and dilatations. Lately this viewpoint has been
recovered~\cite{Goneraetal} using the Hopf dual (see references
therein) of the twist deformation of the Poincar\'e enveloping
algebra.

The twist approach to Poincar\'e spacetime transformations in noncommutative
field theory has its origin in ref.~\cite{Oeckl} and was developed in
ref.~\cite{Kulish}, see also~\cite{Wess}. The authors of ref.~\cite{Kulish}
argue that, if Poincar\'e invariance did not survive in noncommutative field
theory, Wigner's particle classification in terms of scalar, vector, spinor
fields and so on would not be appropriate. An extension to 4-dimensional
special conformal transformations has also been considered~\cite{Matlock}. The
same ideas have been used to study the action of the conformal group in two
noncommutative spacetime dimensions~\cite{Lizzi-Vaidya-Vitale}, where there is
an infinite number of generators. They have also been used to explore
noncommutative formulations of gravity~\cite{Aschieri} and gauge
theories~\cite{twisted-gauge}.
%, whereupon at present the situation is however far from settled.

In this paper we undertake the study of noncommutative spacetime
transformations from the observer point of view with new methods. We
start by considering linear affine transformations and find a
transformation law for~$\Th$ that ensures affine covariance of the
Moyal product. From the infinitesimal formulation of this covariance,
expressions for the transformation generators as differential
operators in an $(x,\Th)$-space are obtained. These generators by
construction satisfy the Leibniz rule for the Moyal product and the
same commutation rules as for commutative spacetime. In terms of them
the study of symmetries becomes very simple, and the discussion of the
invariance of a classical action in field theory for $\Th=0$ is easily
transferred to the case~$\Th\neq 0$. We also consider the
generalization of the twist-deformed enveloping algebra approach to
diffeomorphisms and give a closed form for the twisted coproduct of an
arbitrary infinitesimal generator. This generalization leads to a neat
interpretation of the twist description and allows us a clear
comparison between the covariant description of hidden symmetries and
the twist formalism.

The paper is organized as follows. In Section~2 we show that the Moyal
product is globally covariant under linear affine transformations and
obtain their infinitesimal generators in~\hbox{$(x,\Th)$-space.} We
particularize to Weyl transformations in Section~3 and obtain in
\hbox{$(x,\Th)$-space} their generators
$\{P^\Th_\m,M^\Th_{\m\n},D^\Th\}$. From this, we give a very simple
and straightforward proof of Weyl invariance for noncommutative $U(n)$
gauge theory. Section~4 is dedicated to investigate spacetime
transformations whose infinitesimal form is polynomial in the
Cartesian coordinates, of which special conformal transformations are
an example. We find that the Moyal product is not covariant for them.
In Section 5 we thoroughly compare the observer approach with the
twisted coproduct approach for the Poincar\'e algebra. Section 6
collects our conclusions. We include a short mathematical appendix.

\section{\label{sec:Affine}Linear affine covariance of the Moyal product}

We find in Subsection 2.1 the transformation law for~$\Th$ that
renders the Moyal product covariant under linear affine
transformations of spacetime. In Subsection~2.2 we exhibit the
infinitesimal generators for such transformations.

\subsection{\label{sec:Global}Global analysis}

\indent Given two functions $f$ and $g$ defined on~$\R^4$ and any
$4\x4$ real anti-symmetric matrix~$\Th$, Rieffel's
formula~\cite{Rieffel} for the Moyal star product of~$f$ with~$g$ is
\begin{equation}
   (f \star_\Th g)\,(x) = \frac{1}{(2\pi)^4} \int d^4\!u \,d^4\!v\,
     f\bigl(x + \thalf\Th u\bigr)\,g(x + v)\,e^{iu\. v}\,.
\label{eq:Rieffel-prod}
\end{equation}
This definition generalizes others in the literature in that it does
not require~$\Th$ to be non-degenerate. It is also valid for any
number~$n$ of dimensions, even or odd, by simply replacing~$4$ by~$n$.
With $f=x^\mu$ and $g=x^\nu$, expression~\eqref{eq:Rieffel-prod}
reproduces the commutation relations
\begin{equation*}
        [x^\m, x^\n]_{\star_\Th} := x^\m{\star_\Th}x^\n -
        x^\n{\star_\Th}x^\m = i\th^{\m\n}\,.
\end{equation*}
Formal expansion of $f$ in the integrand in powers of~$\Th$ and
integration by parts yields
\begin{equation}
  (f \star_\Th g)\,(x) =
    {\rm exp}\left[\tihalf\,\th^{\m\n}\>\pa^{(x)}_\m\>\pa^{(y)}_\n\right]
       f(x)\,g(y) \bigg\vert_{y=x}\,,
\label{eq:power-series}
\end{equation}
which is the form for the Moyal product most often quoted in the
literature of field theory. A~rigorous mathematical derivation of a
formula of the type of~\eqref{eq:power-series} as an asymptotic
development of the exact formula can be found in~\cite{Estrada}. It is
important to note, however, that whereas expansions
like~\eqref{eq:power-series} are local,
expression~\eqref{eq:Rieffel-prod} is not.

Let us now consider the group of linear affine transformations. We
recall that a transformation $\Om=(L,a)$ of this type is characterized
by a real $4\x4$ matrix $L$ with nonvanishing determinant and a vector
$a$ in $\R^4$. On a vector $x\in\R^4$ it acts by
\begin{equation}
   x\mapsto \Om\.x =Lx+a\,.
\label{eq:on-x}
\end{equation}
The group product is given by $\Om\Om'= (LL',La' + a)$ and the inverse
of the transformation~$\Om=(L,a)$ is~$\Om^{-1}\!=\!(L^{-1},-L^{-1}a)$.
The action of linear affine transformations on functions on~$\R^4$ is
given by
\begin{equation}
        [\Om\. f]\, (x) = f\bigl(\Om^{-1}\.x\bigr) =
        f\bigl(L^{-1}(x-a)\bigr)\,.
\label{eq:on-f}
\end{equation}
With this definition,  we have $[\Om\. f](\Om\.x)= f(x)$ and
\begin{equation*}
    \Om_1\.[\Om_2\. f]\ = (\Om_1 \Om_2) \. f\,.
\end{equation*}
Had we taken the action on a function~$f$ to be defined by $[\Om\.
f](x)=f(\Om\.x)$, we would have obtained $\Om_1\!\.[\Om_2\.  f]=
(\Om_2\,\Om_1)\.f$, which looks less natural. Here we stick
to~\eqref{eq:on-f}.

To investigate the covariance of the star product $\star_\Th$ under
the linear affine group, we need to compute $[\Om\. f]\star_\Th [\Om\.
g]$. Using Rieffel's definition, we obtain
\begin{equation*}
          \big([\Om\. f] \star_\Th [\Om\. g]\big)\,(x) =
          \frac{1}{(2\pi)^4} \int d^4\!u \,d^4\!v\, f\big(\Om^{-1}\.(x +
          \thalf\Th u)\big)\,g\big(\Om^{-1}\.(x + v)\big) \,e^{iu\. v}\,.
\end{equation*}
Noting that $\,\Om^{-1}\!\.\!(x+x_0)=\Om^{-1}\!\.\!x+L^{-1}x_0\,$ and
making the changes of variables~\hbox{$u\to (L^{-t})\,u$}, where
$L^{-t}:=(L^{-1})^t=(L^t)^{-1}$ is the contragredient matrix, and
\hbox{$v\to Lv$}, we arrive at
\begin{equation*}
   \big([\Om\. f] \star_\Th [\Om\. g]\big)\,(x) =
      \frac{1}{(2\pi)^4} \int d^4\!u \,d^4\!v\,
      f\big(\Om^{-1}\.x + \thalf L^{-1}\Th L^{-t} u\big)\,
       g\big(\Om^{-1}\.x + v\big) \,e^{iu\. v}\,.
\end{equation*}
It is then clear that if the action of a linear affine transformations
on the space of anti-symmetric matrices $\Th$ is defined as
congruence,
\begin{equation}
        \Om\.\Th = L\Th L^t\,,
\label{eq:on-theta}
\end{equation}
one has
\begin{equation}
  [\Om\. f] \star_\Th [\Om\. g]
       = \Om \. \big( f \star_{\Om^{-1}\.\Th} g\big)
  \qquad{\rm or} \qquad
  [\Om\. f] \star_{\Om\.\Th} [\Om\. g] = \Om\.(f \star_\Th g)\,.
\label{eq:covariance}
\end{equation}
Identities~\eqref{eq:on-x} to~\eqref{eq:covariance} constitute the
starting point for our analysis. They show that the Moyal product is
fully covariant under linear affine transformations, provided the
matrix $\Th$ transforms as in~\eqref{eq:on-theta}. One could say that
we have generalized `observer' covariance to linear affine
transformations. For transformations $\Om$ such that $\Om\.\Th=\Th$,
equations~\eqref{eq:covariance} take the form
\begin{equation}
     [\Om\. f] \star_\Th [\Om\. g] = \Om\. (f \star_\Th g)\,.
\label{eq:particle}
\end{equation}
In this case there is no distinction between `observer' and `particle'
linear affine transformations.

Let us study in some detail the set of transformations $\Om$ for which
$\Om\.\Th=\Th$. Since the only invariant of congruence is the rank,
and any real anti-symmetric matrix $\Th$ has rank 4, 2 or 0, there are
only three orbits of the action~\eqref{eq:on-theta}. They respectively
correspond to the generic set of invertible anti-symmetric matrices,
to the set of non-invertible, nonvanishing anti-symmetric matrices,
and to the zero matrix. Assume that $\Th$ is in the rank-4 orbit and
consider matrices $L$ such that $L\Th L^t=\Th$. They form a group,
that we may call the little group. Since the dimension of the little
group is that of its Lie algebra, to find it suffices to use the
exponential form $L=e^B$ and require $L\Th L^t=\Th$ to first order in
$B$. This yields the condition $B\Th=(B\Th)^t$, which in turn gives
ten independent entries for $B$. Hence the little group $G_4$ of
linear affine transformations that leave invariant the rank-4 orbit
has dimension 14, counting the four translational degree of freedom.
Similar arguments show that the little group $G_2$ of the rank-2 orbit
has dimension 15.

Even if the functions $f$ and $g$ do not depend explicitly on~$\Th$,
their Moyal product does. On the basis of this fact, and in accordance
with the spirit of covariance, it is convenient to consider an
$(x,\Th)$-space on which a linear affine transformation acts as
\begin{equation}
   \Om\.(x,\Th) = (L,a)\.(x,\Th) = (Lx+a,L\Th L^t)\,.
\label{eq:on-x-theta}
\end{equation}
We emphasize that the coordinates which parametrize the variables~$x$
and~$\Th$ both change under such a transformation.

\subsection{\label{sec:Infinitesimal}Infinitesimal generators}

The action \eqref{eq:on-x-theta} possesses infinitesimal generators,
which we will generically denote $G^\Th$ and which are vector fields
in $(x,\Th)$-space. As convenient coordinates to express them, we may
choose the spacetime coordinates $x^\m$ and six independent entries
$\th^{\m\n}$ of $\Th$. We have for small~$B$
\begin{equation*}
    \Om\.(x,\Th) = \bigl(x + Bx + a,\Th + B\Th + \Th B^t\bigr)
    + O(B^2)\,.
\end{equation*}
From this we read the generators, when acting on functions, as
\begin{equation}
  G^\Th := -\bigl( a^\a + B^\a_{\,\b}\> x^\b \bigr)\, \frac{\pa}{\pa x^\a}
           - \frac{1}{2}\, \bigl( B^\a_{~\ga}\, \th^{\ga\b}
              + \th^{\a\ga}\, B^\b_{~\ga}\bigr)\>
\frac{\pa}{\pa\th^{\a\b}}\,,
\label{eq:preliminary}
\end{equation}
where we have put $B=(B^\a_{\,\b})$. The factor~$\thalf$ in front of
the last parenthesis arises because we have chosen as coordinates for
$\Th$ its entries with, say, $\a<\b$ and in~\eqref{eq:preliminary} we
are summing over all~$\a,\b$. We may recast~\eqref{eq:preliminary} as
\begin{equation}
    G^\Th = G^\Th_a + G^\Th_B\,,
\label{eq:generator}
\end{equation}
where
\begin{equation}
  G^\Th_a = -a^\a \pa_\a  \qquad
  G^\Th_B = -\,\eps^\a\,\pa_\a + \frac{1}{2}\, \dl_\eps\th^{\a\b}\>
\frac{\pa}{\pa\th^{\a\b}}\,.
\label{eq:gen}
\end{equation}
Here $\eps(x)$ is the vector field with components $\eps^\a(x) =
B^\a_{\,\b}x^\b$, and
\begin{equation}
 \dl_\eps\,\th^{\a\b} = - \big( B^\a_{~\ga}\, \th^{\ga\b}
    + \th^{\a\ga}\, B^\b_{~\ga} \big)
\label{eq:Lie-th}
\end{equation}
are the components of the Lie derivative with respect to~$\eps(x)$ of
the 2-tensor $\,\Th\!=\!\th^{\a\b}\pa_\a \ox\pa_\b$. Here we choose
not to distinguish the matrix~$\Th$ and the corresponding tensor in
the notation. We recall in this regard that the Lie derivative with
respect to a vector field $\,v(x)\!=v^\ga(x)\,\pa_\ga$ of a
contravariant 2-tensor with components~$t^{\rho\sigma}$ is given by
\begin{equation*}
   \dl_v t^{\a\b}  = v^\ga\,\pa_\ga t^{\a\b}
         - t^{\ga\b}\pa_\ga v^\a - t^{\a\ga}\pa_\ga v^\b\,,
\end{equation*}
which reduces to~\eqref{eq:Lie-th} for $v^\a=\eps^\a$
and~$t^{\a\b}\!=\!\th^{\a\b}$ independent of~$x$. In $G^\Th_a$ we
recognize the generators of translations. The term $-\eps^\a\pa_\a$
in~$G^\Th_B$ generates arbitrary linear spacetime transformations.
Finally, the term $\thalf\dl_\eps\th^{\a\b}\,\pa/\pa\th^{\a\b}$
accounts for the linear transformations in the $\Th$-directions that
ensure covariance. The action of the operator~$G^\Th$ on the Moyal
product $f\!\star_\Th g$ directly follows from the covariance
law~\eqref{eq:covariance}. Indeed, the infinitesimal version of the
latter simply states that
\begin{equation}
     G^\Th(f \star_\Th g) = G^\Th\! f \star_\Th g + f \star_\Th G^\Th g\,.
\label{eq:Leibniz}
\end{equation}
Hence the generators $G^\Th$ do satisfy the Leibniz rule for the Moyal
product of functions. In mathematical terms, the $G^\Th$ are
derivations of the Moyal algebra. Note that our way to proceed, i.e.,
descending from global to infinitesimal covariance, identifies the
generators and establishes that they are derivations all at once.

It remains to make explicit the action of each of the generators
in~\eqref{eq:gen} on the Moyal product. To do this, we use again
Rieffel's formula~\eqref{eq:Rieffel-prod} and obtain, after some
algebra,
\begin{eqnarray}
  & {\displaystyle
    \pa_\a (f \star_\Th g) =  \pa_\a f \star_\Th g
       +  f \star_\Th \pa_\a g
    } & \label{eq:translation} \\
  & {\displaystyle
    x^\a ( f\star_\Th g) = x^\a f\star_\Th g
       - \frac{i}{2}\>\th^{\a\b} f\star_\Th \pa_\b g
                       =  f\star_\Th x^\a  g
       + \frac{i}{2}\>\th^{\a\b}\, \pa_\b f \star_\Th g
    } & \label{eq:non-derivation}\\
  & {\displaystyle
     \frac{\pa}{\pa\th^{\a\b}}\> (f \star_\Th g)
         = \frac{\pa f}{\pa\th^{\a\b}} \star_\Th g
         + f \star_\Th \frac{\pa g}{\pa\th^{\a\b}}
         + \frac{i}{2}\> (\pa_\a f \star_\Th \pa_\b g
               - \pa_\b f \star_\Th \pa_\a g)\,.
    } & \label{eq:rare}
\end{eqnarray}
Eqs.~\eqref{eq:translation} and~\eqref{eq:non-derivation}
together imply that
\begin{equation}
  \eps^\a \pa_\a ( f\star_\Th g) =  \eps^\a \pa_\a f\star_\Th g
     + f \star_\Th \eps^\a \pa_\a g
     + \frac{i}{2}\> \dl\th_{\eps}^{\a\b} \pa_\a f \star_\Th \pa_\b g\,.
\label{eq:order-1}
\end{equation}
These calculations serve as a check of our argument. Indeed, it is
straightforward to see from eq.~\eqref{eq:gen} that
eqs.~\eqref{eq:rare} and~\eqref{eq:order-1} imply the Leibniz
rule~\eqref{eq:Leibniz}. Particularly, for~$f$ and~$g$ independent
of~$\Th$, one has the elegant
\begin{equation}
   -G^\Th_B(f \star_\Th g) = \eps f \star_\Th g + f \star_\Th \eps
         g\,,
\label{eq:particular}
\end{equation}
where still $\eps=B^\a_\b x^\b\pa_\a$. The previous formulae are the
heart of the paper, and many things in it flow from them.

In summary, the generators of translations are derivations, and the
generators of linear affine transformations can be made into
derivations if such transformations are accompanied by suitable
transformations in the $\Th$-directions. Eq.~\eqref{eq:Leibniz} tell
us that vector fields with components of degree up to one in the
coordinates $x^\m$ can still be regarded as generating symmetries of
the Moyal algebra. For this, the linearity in the coordinates $x^\m$
of the components of the vector field $\eps(x)$ is decisive. In
Section~4 we show that matters are different for higher order
polynomial dependence. Note also that, for the simpler case of~$\R^2$,
the matrix~$\Th$ has components $\th^{\a\b}= \th\epsilon^{\a\b}$. If
in addition $B^\a_{\,\b}=-\dl^\a_{\,\b}$ ---the sole nontrivial
possibility--- we recover the derivation $G=x\!\.\!\pa+
2\th\frac{\pa}{\pa\th}$, already obtained in~\cite{Gutt}. From this
point of view, our results can be seen as an expansion of the scenario
of that paper. See the Appendix for further discussion of this point.

\section{\label{sec:Weyl}The Weyl group on $\boldsymbol{(x,\Th)}$-space and
  its action on fields}

So far nothing we have said involves a choice of metric, nor of an
action. Approaches to physical problems based on the affine group are
known, the metric-affine theory of gravity~\cite{Hehl} among them.
However, conventional physical theories require the choice of a
metric. Since we are ultimately interested in field theory on
Minkowski spacetime, let us adopt the Minkowskian metric. Everything
works much the same for the Euclidean metric. Furthermore, we consider
the Weyl group~$W$ of translations, Lorentz transformations and
dilatations. It is a subgroup of the linear affine group and hence the
results of the last section apply. According to them, the generators
of Lorentz transformations and dilatations in $x$-space receive
contributions involving derivatives with respect to~$\th^{\m\n}$. In
what follows we describe the algebra of these generators.

\subsection{\label{sec:Algebra}The Weyl algebra in
  $\boldsymbol{(x,\Th)}$-space.} 

In the presence of a metric we are allowed to lower and raise the
indices of the different tensors. If~$L$ in eq.~\eqref{eq:on-x-theta}
accounts for a Lorentz transformation $\dl x^\a=\om^{\a\b}x_\b-
\om^{\b\a}x_\b$, we can take for the matrix~$B$ in
equation~\eqref{eq:preliminary} any of the matrices $M_{\m\n}$ with
entries $(M_{\m\n})^\a_{~\b}= \dl_{~\m}^\a\,g_{\n\b}-
\dl_{~\n}^\a\,g_{\m\b}$. Substituting in formula~\eqref{eq:gen}, we
obtain for the derivations in~$(x,\Th)$-space associated to Lorentz
transformations
\begin{equation}
                M^\Th_{\m\n} = x_\m\pa_\n - x_\n\pa_\m +
                \th^{\rho}_{~\m}\frac{\pa}{\pa\th^{\rho\n}} -
                \th^{\rho}_{~\n}\frac{\pa}{\pa\th^{\rho\m}}\,.
\label{eq:gen-Lorentz}
\end{equation}
Analogously, if $L$ accounts for a dilatation $\dl x^\m=\la x^\m\,$,
formula~\eqref{eq:Lie-th} yields~$\,\dl\th^{\m\n}= -2\la\theta^{\m\n}$
and for the the corresponding derivation we have
\begin{equation}
          D^\Th = -x\!\.\!\pa - \th^{\m\n}\frac{\pa}{\pa\th^{\m\n}}~.
\label{eq:gen-dilatations}
\end{equation}
By construction, the operators~$M^\Th_{\m\n}\>,~D^\Th$ and the
generators $P^\Th_\m=-\pa_\m$ of translations are derivations of the
Moyal star product. It is easy to check that they satisfy the same
commutation relations as the generators of the ordinary Weyl Lie
algebra, namely
\begin{equation}
\begin{array}{c}
     [P^\Th_\mu, P^\Th_\nu] = 0 \qquad
     [P^\Th_\mu, M^\Th_{\n\rho}]
               = g_{\m\n}P_\rho^\Th - g_{\m\rho} P^\Th_\n \qquad
     [D^\Th,P^\Th_\m] = P^\Th_\m \smallskip \cr
     [D^\Th, M^\Th_{\m\n}] = 0 \qquad
     [M^\Th_{\m\n}, M^\Th_{\rho\sigma}] = g_{\n\rho}\,M^\Th_{\m\sigma}
          + g_{\m\sigma}\,M^\Th_{\n\rho} - g_{\m\rho}\,M^\Th_{\n\sigma}
          - g_{\n\sigma}\,M^\Th_{\m\rho}\,.
\end{array}
\label{eq:commutators}
\end{equation}
We conclude that $\{P^\Th_\m,M^\Th_{\m\n},D^\Th\}$ represent the Weyl
Lie algebra in~$(x,\Th)$-space.

\subsection{\label{sec:Actions}Weyl invariant actions.}

In field theory we are interested in fields. We expect them to
transform according to irreducible representations of the Poincar\'e
Lie algebra. Let us consider for example a local $U(1)$-$\star$ gauge
field $A_\m(x)$, with classical action
\begin{equation}
     S[A] = -\frac{1}{4g^2}\int d^4\!x ~ F^\Th_{\m\n}(x)
         \star F^{\Th\,\m\n}(x)\,,
\label{eq:U(1)-action}
\end{equation}
where
\begin{equation}
  F^\Th_{\m\n} = \pa_\m A_\n - \pa_\n A_\m - i\big( A_\m\star_\Th A_\n -
             A_\n\star_\Th A_\m \big)
\label{eq:field-strength}
\end{equation}
denotes the noncommutative field strength. We recall that this action
is invariant under noncommutative $U(1)$~local gauge transformations,
whose infinitesimal form is~$\dl A_\m=\pa_\m\la-i(A_\m\star_\Th\la
-\la\star_\Th A_\m)$.

Since $A_\m(x)$ does not depend on $\Th$, its variation under an
infinitesimal transformation
\begin{equation*}
 \dl A_\a(x) =  [\Om\!\.\!A_\a](x) - A_\a(x)\, ,
\end{equation*}
which in field theory is written as $\dl A_\a(x) = A'_\a(x) -
A_\a(x)$, is the same as for $\Th=0$. This means that the action of
the generators $\{ P^\Th_\m, M^\Th_{\m\n}, D^\Th\}$ on $A_\a(x)$ is
the usual one,
\begin{align*}
    P_\m^\Th[A_\a] &= -\pa_\m A_\a \\
    M_{\m\n}^{\Th}[A_\a] &= (x_\m\pa_\n - x_\n\pa_\m)A_\a + g_{\m\a}A_\n
                          - g_{\a\n}A_\m \\
     D^{\Th}[A_\a] & = -\,(1 + x\!\.\!\pa)A_\a\,.
\end{align*}
The key point now is that, since $P^\Th_\m,\>M^\Th_{\m\n},\>D^\Th$ are
derivations for the Moyal star product, their action on the field strength
$F^\Th_{\m\n}$ is functionally the same for all~$\Th$, and in particular equal
to that for~$\Th=0$. This automatically leads to
\begin{align*}
    P^\Th_\a[F^\Th_{\mu\nu}] & = \pa_\a F^\Th_{\mu\nu} \\
    M^\Th_{\a\b}[F^\Th_{\mu\nu}] & = x_\a\pa_\b F^\Th_{\mu\nu}
     - x_\b\pa_\a F^\Th_{\mu\nu}
     + g_{\mu\a}F^\Th_{\b\nu} - g_{\mu\b}F^\Th_{\a\nu}
     + g_{\nu\a}F^\Th_{\b\mu} - g_{\nu\b}F^\Th_{\a\mu}  \\
    D^\Th[F^\Th_{\mu\nu}] & = -\,( 2 + x\!\.\!\pa)\, F^\Th_{\mu\nu}\,.
\end{align*}
The proof of Weyl invariance of $S[A]$ then goes as in the commutative case.
We thus recover, without recourse to functional derivatives, the results of
ref.~\cite{Vienna}. With obvious changes, the arguments above apply to
noncommutative $U(n)$ gauge fields. In conclusion, covariance of the Moyal
product, hence the knowledge of generators which are derivations of the Moyal
product, simplifies the proof of invariance of a field theory action.

\section{\label{sec:Nogo}Non-affine spacetime transformations}

The question that naturally arises is whether covariance can be extended to
spacetime transformation groups whose generators have coefficients with
arbitrary polynomial dependence on the coordinates~$x^\m$. We answer this
question in the negative.

Let us consider a spacetime transformation $\,x\to\Om\.x\,$ whose
infinitesimal form is quadratic in~$x$. It is generated by linear combinations
of differential operators of the form $x_\m x_\n \pa_\rho$. For the Moyal
product to be covariant under such transformation, the matrix $\Th$ must
transform in such a way that the infinitesimal form~\eqref{eq:Leibniz} of
covariance holds. This requires the existence of sets of generators, say
$\{G^\Th_{\m\n\rho}\}$, defined as differential operators in
\hbox{$(x,\Th)$-space} which act on the Moyal product as derivations and
reduce to~$\{\!\,x_\m x_\n \pa_\rho\,\!\}$ for $\Th=0$. Now, acting with~$x_\m
x_\n \pa_\rho$ on $\,f\star_\Th g\,$ and using
equations~\eqref{eq:non-derivation} and~\eqref{eq:rare}, we obtain
\begin{align*}
   x_\m x_\n \pa_\rho (f\star_\Th g) & =  x_\m x_\n \pa_\rho f\star_\Th g
     + f\star_\Th  x_\m x_\n \pa_\rho g  \nonumber \\
  & - \left( \th^\a_{~\m} x_\n + \th^\a_{~\n} x_\m \right)\,
      \Big[ \frac{\pa}{\pa\th^{\a\rho}}(f\star_\Th g)
          - \frac{\pa f}{\pa\th^{\a\rho}} \star_\Th g
          - f\star_\Th \frac{\pa g}{\pa\th^{\a\rho}} \Big] \nonumber\\
  & + \frac{1}{4}\, \th^\a_{~\m} \th^\b_{~\n} \,
      \Big( \pa_\rho f \star_\Th \pa_\a\pa_\b g
       + \pa_\a\pa_\b f  \star_\Th \pa_\rho g \Big)\,.
\end{align*}
It follows that the operator
\begin{equation*}
  G^\Th_{\m\n\rho}=x_\m x_\n\pa_\rho+
     \left(\th^\a_{~\m}x_\n+\th^\a_{~\n}x_\m\right)\,
        \frac{\pa}{\pa\th^{\a\rho}}
\end{equation*}
satisfies
\begin{equation}
  G^\Th_{\m\n\rho}  (f\star_\Th g)
   =  G^\Th_{\m\n\rho}f\star_\Th g
   + f \star_\Th  G^\Th_{\m\n\rho} g
   + \frac{1}{4} \, \th^{\m\a} \th^{\n\b} \,
      \Big( \pa_\rho f \star_\Th \pa_\a\pa_\b g
       + \pa_\a\pa_\b f  \star_\Th \pa_\rho g \Big)\,.
\label{eq:almost}
\end{equation}
The last two terms in this equation, containing three partial derivatives,
cannot be recast as a derivation for $\star_\Th$ and thus prevent the operator
$G^\Th_{\m\n\rho}$ from being a derivation.  Therefore we do not see a way for
the operators~$x_\m x_\n\pa_\rho$ to be made into derivations in
$(x,\Th)$-space so that the spacetime transformations that they generate
become a covariance of the Moyal product.

In particular, special conformal transformations cannot become a covariance of
the Moyal product, since they are generated by vector
fields~\hbox{$K_\m=x^2\pa_\m-2x_\m x\!\.\!\pa$}, which are particular linear
combinations of~$x_\m x_\n \pa_\rho$. The same clearly holds for vector fields
with higher order polynomial dependence on~$x^\m$.

\section{\label{sec:Comparison}Comparison with the twist-deformed enveloping
  algebra formalism}
  
Our aim here is to understand the twist-deformed description of spacetime
transformations for the Moyal product in the light of our approach. We start
by observing that the twist deformation of the Poincar\'e enveloping algebra
used in~\cite{Kulish} to describe noncommutative spacetime and its
transformations is a particular instance of a well known procedure to `twist'
Hopf algebras, originally due to Drinfeld. See ref.~\cite{Majid} for a review
and details. In what follows we briefly recall it. If $H$ is a Hopf algebra,
denote by $\,\id\,$ the identity map of~$H$ onto itself, by $\Dl$ the
coproduct map, and by $\eta$ the counit map from the Hopf algebra to the
scalars. Consider an invertible element $\chi$ in $H\ox H$ that satisfies the
conditions

\begin{equation}
   (1\ox\chi)(\id\ox\Dl)\chi = (\chi\ox1)(\Dl\ox\id)\chi
   \qquad\qquad
   (\eta\ox\id)\chi = (\id\ox\eta)\chi = 1\,.
\label{eq:coboundary}
\end{equation}
The element $\chi$ is said to be a counital 2-cocycle for~$H$. For such a
$\chi$, the twist $\Dl_\chi(h) =\chi\Dl(h)\chi^{-1}$, with $h$ in $H$, does
define a new coproduct in $H$. The algebra underlying $H$ endowed with the new
coproduct $\Dl_\chi$ is still a Hopf algebra, called twisted Hopf algebra,
which may be denoted by~$H_\chi$. As Hopf algebras, $H_\chi$ and~$H$ are
isomorphic if $\chi$ has the trivial form $\chi= (\ga\ox \ga)\Dl\ga^{-1}$,
with $\ga$ an invertible element in $H$ satisfying $\eta(\ga)=1$. Assume
moreover that $H$ has a representation in an associative algebra~$\F$ with
product $m$. That is, for~$h$ in~$H$ and $a,b$ in~$\F$ one has
\begin{eqnarray}
   & m(a\ox b) = ab & \label{eq:m} \\[3pt]
   & {\displaystyle
   h\.(ab) = h\. m\,(a\ox b) = m \big(\Dl(h)\.(a\ox b)\big)\,, } &
\label{eq:m-covariance}
\end{eqnarray}
where $m(a,b)$ denotes the product of~$a$ and~$b$ in~$\F$. The twisting
of~$\Dl$ introduces in~$\F$ a twisted product $m_\chi$ defined by
\begin{equation}
      m_\chi(a\ox b) = m\big(\chi^{-1}\.(a\ox b)\big)\,.
       \label{eq:m-chi}
\end{equation}
$H_\chi$ is represented in the new algebra by its action through
$\Dl_\chi(h)$, since
\begin{align}
     h\.m_\chi(a\ox b) & = h\. m\big(\chi^{-1}\.(a\ox b)\big) \nonumber\\
       & = m\big(\Dl(h)\chi^{-1}\.(a\ox b)\big)  \nonumber\\
       & = m\big(\chi^{-1}\Dl_\chi(h)\.(a\ox b)\big)  \nonumber\\
       & = m_\chi\big(\Dl_\chi(h)\.(a\ox b)\big)\,.
          \label{eq:twist-covariance}
\end{align}
Here we have used eq.~\eqref{eq:m-chi}, eq.~\eqref{eq:m-covariance},
the definition of $\Dl_\chi$ and eq.~\eqref{eq:m-chi} in this order.
This equation will play a central part below in understanding how the
Moyal product behaves under general spacetime transformations.
Furthermore, it is easy to see that the first condition
\eqref{eq:coboundary} implies that the new twisted product $m_\chi$ is
associative.

Now, let us consider the Lie algebra $\diff$ of diffeomorphisms, whose
generators are vector fields with polynomial coefficients on~$\R^4$.
As Hopf algebra $H$ we take the enveloping algebra~$\U(\diff)$.
Likewise the enveloping algebra of any Lie algebra, the coproduct
$\Dl$ is first defined for elements $h$ of $\diff$ by $\Dl(h)=1\ox
h+h\ox1$, and then multiplicatively extended to all of $\U(\diff)$ by
means of $\Dl(hh')=\Dl(h)\Dl(h')$. For the algebra $\F$ carrying a
representation of $\U(\diff)$, take the algebra of functions on
spacetime with the ordinary multiplication $m(f\ox g)=fg$. Finally,
for $\chi$, we take the exponent of the Poisson tensor
$\chi_\Th=\exp(-\tihalf\,\th^{\m\n}\pa_\m\ox\pa_\n )$. This $\chi_\Th$
is clearly in $\U(\diff)\ox\U(\diff)$, has an inverse
\begin{equation*}
 \chi^{-1}_\Th = \exp(\tihalf\,\th^{\m\n}\pa_\m\ox\pa_\n)
\end{equation*}
and satisfies the cocycle condition nontrivially. The Moyal product
under the asymptotic guise~\eqref{eq:power-series} is then recovered
as the `twisted' product
\begin{equation}
   m_{\chi_\Th}(f\ox g) = m\big( \chi_\Th^{-1} \.(f\ox g)\,\big)
      = f\star_\Th g\,.
\label{eq:twisted-product}
\end{equation}
In view of~\eqref{eq:twist-covariance}, it is clear that the action of
a generator $h$ on the Moyal product is determined
by~$\Dl_{\chi_\Th}(h)$, and conversely. In the sequel, for simplicity
of notation, we write~$\Dl_\Th$ for~$\Dl_{\chi_\Th}$ and~$m_\Th$
for~$m_{\chi_\Th}$. For the generators of translations, Lorentz
transformations~\cite{Kulish} and dilatations~\cite{Matlock} the
following expressions were obtained, in our notation,
\begin{align}
   \Dl_{\Th}(P_\m) & = P_\m\ox 1 + 1\ox P_\m  \nonumber\\[3pt]
   \Dl_{\Th}(M_{\m\n}) & = M_{\m\n} \ox 1 + 1 \ox M_{\m\n}
       \nonumber\\
   & + \tihalf\,\th^{\a\b}
           \big[ (g_{\m\a}P_\n - g_{\n\a}P_\m) \ox P_\b
                + P_\a \ox (g_{\m\b}P_\n -  g_{\n\b}P_\m)\big]
       \label{eq:complicated} \\[3pt]
   \Dl_{\Th}(D) & = D \ox 1 + 1 \ox D - i\,\th^{\m\n} P_\m \ox P_\n\,.
      \nonumber
\end{align}
These equations precisely show that $\Dl_{\Th}(M_{\m\n})$ and
$\Dl_{\Th}(D)$ are not derivations of the Moyal product. From
eq.~\eqref{eq:complicated} it was concluded that the Poincar\'e group
remains relevant in noncommutative field theory. Whereas the argument
is suggestive, it does not directly concern Poincar\'e invariance.
Note that eq.~\eqref{eq:twist-covariance} reflects a rather general
geometrical fact, since it places \textit{no restriction} on the
generator~$h$ except that of being an infinitesimal diffeomorphism.
(The limitation to polynomial coefficients in~$h$ arises because
formula~\eqref{eq:power-series} can only deal with diffeomorphisms of
this type.) This is why the generators $K_\mu$ of special conformal
transformation could be added to the list of
computed~$\Dl_\Th(h)$~\cite{Matlock}. Now, because we are in the
enveloping algebra, eq.~\eqref{eq:twist-covariance} applies to
differential operators of any order. The method is thus a recipe to
encode the action of arbitrary differential operators with polynomial
coefficients on Moyal products.

The previous remark leads in a systematic and simple way to compute
the twisted coproduct of the generator of any spacetime
transformation. Let us take an infinitesimal spacetime transformation
generated by differential operators of the form $x^{\m_1}\cdots
x^{\m_N}\pa_\n$. Using equations~\eqref{eq:translation}
and~\eqref{eq:non-derivation} to compute its action on $f\star_\Th g$,
and invoking the definition of the twisted product $m_\Th$
in~\eqref{eq:m-chi} and its covariance
property~\eqref{eq:twist-covariance}, it follows that
\begin{align}
   \Dl_\Th(x^{\m_1}&\cdots x^{\m_N}\pa_\n) =
        x^{\m_1}\cdots x^{\m_N}\pa_\n \ox 1
      + 1 \ox x^{\m_1}\cdots x^{\m_N}\pa_\n \nonumber\\
   &+ \sum_{k=1}^N \Big(\frac{i}{2}\Big)^k \!\! \sum_{N\ge c_k>\cdots
   >c_1\ge1} \! \th^{\m_{c_1}\a_{c_1}}\, \cdots\,
   \th^{\m_{c_k}\a_{c_k}}
           \Big[\, \pa_{\a_{c_1}}\!\cdots\,\pa_{\a_{c_k}} \ox x^{\m_1}
           \cdots\smile\hspace{-12pt}{}^{c_1}
           \cdots\smile\hspace{-12pt}{}^{c_k}\cdots \,x^{\m_N}\,\pa_\n
           \nonumber\\
   & \hspace{110pt} + (-1)^k\, x^{\m_1}
   \cdots\smile\hspace{-12pt}{}^{c_1}
   \cdots\smile\hspace{-12pt}{}^{c_k}\cdots \,x^{\m_N}\,\pa_\n \ox
   \pa_{\a_{c_1}}\!\cdots\,\pa_{\a_{c_k}} \Big]\,.
\label{eq:general}
\end{align}
Here $\smile\hspace{-12pt}{}^{c_1}$ indicates that in the product
$x^{\m_1}\cdots x^{\m_N}$ the factor $x^{\m_{c_1}}$ is removed. An
equivalent formula can be found in ref.~\cite{Aschieri}.
With~\eqref{eq:general}, eqs.~\eqref{eq:complicated} follow at a
stroke. Moreover, it is straightforward to verify that
\begin{equation*}
   m_\Th\big( \Dl_\Th(x^{\m_1}\cdots x^{\m_N}\pa_\n)\.
     (x^\a\ox x^\b - x^\b \ox x^\a)\big) = 0\,.
\end{equation*}
In other words, $\th^{\a\b}$ remains unchanged. The twisted coproduct
formulation accounts only for particle transformations, for transformations
of~$\Th$ are left out. To understand observer transformations, one has to look
elsewhere, to the analysis presented in this paper.

To summarize our comparison, for $G$ in the affine group the relation between
the covariant and twist approaches can be accounted by the following equation
\begin{equation*}
   m_\Th\big( \Dl_\Th(G)\.(f\ox g) \big) = G^\Th m_\Th(f\ox g) 
     - \frac{1}{2}\,\dl_G\th^{\a\b}\, \frac{\pa}{\pa\th^{\a\b}}\,
     m_\Th(f\ox g)\,,
\end{equation*}
where we recall $\dl_G\th^{\a\b}$ is the Lie derivative of the tensor
$\Th=\th^{\a\b}\pa_\a\ox\pa_\b$ with respect to $G$, see
definition~\eqref{eq:preliminary}. For the sake of illustration, this means
that for instance for dilatations one has
\begin{equation*}
   m_\Th\big( \Dl_\Th(D)\.(f\ox g) \big) = D^\Th (f\star_\Th g) 
     + \th^{\a\b}\,\frac{\pa}{\pa\th^{\a\b}}\,(f\star_\Th g) \,.
\end{equation*}
Furthermore, observer and twist covariances boil down to
\begin{equation*}
    {\rm observer}\!:~G^\Th m_\Th = m_\Th\Dl(G)
    \qquad
    {\rm twist}\!:~G\, m_\Th = m_\Th\Dl_\Th(G)\,.
\end{equation*}

\section{\label{sec:Conclusion}Conclusion and outlook}

We have investigated noncommutative spacetime transformations from the
observer point of view. This regards transformations as coordinate changes
under which both the fields and the noncommutativity matrix tensor $\Th$
transform. The dependence of the Moyal product $f\star_\Th g$ on the spacetime
point $x$ and the matrix $\Th$ leads naturally to introduce an
$(x,\Th)$-space. This is in the spirit of the widespread belief that
noncommutativity arises at a certain fundamental length, let it be the Planck
length or other. Our main result is that the Moyal product is covariant under
linear affine transformations. We obtain explicit expressions for the
generators of the Weyl transformations in $(x,\Th)$-space satisfying two very
important properties: they are derivations for the Moyal product and represent
the Weyl Lie algebra.\footnote{Expressions for the $\th^{\a\b}\!$-derivative
  terms in these generators have been searched for in the past. See M.
  Chaichian, K. Nishijima and A. Tureanu, ``An interpretation of non
  commutative field theory in terms of a quantum shift '', Phys. Lett. {\bf
    B633} (2006) 129 [arXiv:hep-th/0511094] for an approach to find what we
  now understand to be $M^\Th_{\m\n}-M_{\m\n}$.}.  This strikingly simplifies
the analysis of symmetries of noncommutative field theory actions.  The twist
approach to noncommutative spacetime transformations has also been revisited
and generalized. It is important to remark in connection with the twist
formulation, or for that matter, with any formulation of spacetime
transformations, that knowledge of a family of generators by itself does not
imply invariance of a noncommutative field theory action. This is something
that remains to be elucidated.

Possible venues for the future include the understanding of noncommutative
gauge transformations in terms of covariance, for a mixing of spacetime
coordinates, $\Th$-variables and gauge degrees of freedom takes
place~\cite{Lizzi-Szabo-Zampini}. One may investigate, somewhat along the
lines of~\cite{Vienna,Stern}, the connection of our approach with the
Seiberg--Witten map~\cite{Seiberg-Witten}. It would also be interesting to
study spacetime transformations in which the transformed $x^{\prime\m}$
depends on both $x^\m$ and $\th^{\a\b}$.  Finally, it is worth extending the
covariant approach here to non(anti)commutative superspace~\cite{Seiberg}. The
latter makes sense, since having a representation of the Weyl Lie algebra, it
is sensible to ask for supersymmetric extensions in superspace.\footnote{For
  the twist approach, this was first considered in M. Ihl and C. Saemann,
``Drinfeld-twisted supersymmetry and non-anticommutative superspace'',
JHEP {\bf O601} (2006) 065 [arXiv:hep-th/0506057].
%%CITATION = HEP-TH 0506057;%%
See also Banerjee, C. Lee and S. Siwach, ``Deformed conformal and
super-Poincar\'e symmetries in the non-(anticommutative) space'',
[hep-th/0511205].
%%CITATION = HEP-TH 0511205;%%
}

\begin{acknowledgments}
  The authors are grateful to V.~Gayral, G.~Marmo and C.~Moreno for
  discussions. JMG-B acknowledges support from MEC, Spain, through a `Ram\'on
  y Cajal' contract. Partial support from CICyT and UCM-CAM, Spain, through
  grants~FIS2005-02309,~910770, and from the `Progetto di Ricerca di Interesse
  Nazionale', Italy, is also acknowledged.
\end{acknowledgments}

\appendix

\section{On derivations of the Moyal algebra.}

The purpose of this mathematical note is to prove existence of outer
derivations for a Moyal algebra. This lies outside the main line of the paper,
but stems from the considerations in Section~2. Being a relevant point that
seems to run against standard lore~\cite{Maeda}, we include it here. A
derivation of an associative algebra~$\F$ is a linear map $D:\F\to\F$
satisfying the Leibniz rule
\begin{equation*}
  D(ab) = Da\,b + aDb\,.
\end{equation*}
We convene in saying that a derivation~$D$ is inner if there is an
element $a_D$ in~$\F$ or in a multiplier algebra of~$\F$ such that
\begin{equation*}
  Db = a_Db - ba_D\,.
\end{equation*}
It is said to be outer if not of this form. For simplicity, consider
the Moyal plane $\R^2$. We now follow~\cite{Maeda} inasmuch as
possible. A derivation of the Moyal algebra is entirely determined by
its action on~$x^1,\,x^2$. Let $D_1,\,D_2$ be two derivations with
\begin{equation*}
   D_1x^1 = D_2x^1 = f_1(x^1,x^2) \quad{\rm and}\quad
   D_1x^2 = D_2x^2 = f_2(x^1,x^2)\,.
\end{equation*}
The difference $D_1-D_2$ is another derivation. It is easily checked to
annihilate the plane waves
\begin{equation*}
  (D_1-D_2)e^{i(\a x^1 + i\b x^2)} = 0\,.
\end{equation*}
It then vanishes. Let now~$D$ be an arbitrary derivation. For $D$ to
be inner a function~$a_D$ must exist such that
\begin{equation*}
   [a_D, x^1]_{\star_\Th} = -i\th\,\frac{\pa a_D}{\pa x^2} = f_1
   \qquad \quad
   [a_D, x^2]_{\star_\Th} = i\th\,\frac{\pa a_D}{\pa x^1} = f_2.
\end{equation*}
This is the case if and only if the integrability condition $\pa
f_1/\pa x^1+\pa f_2/\pa x^2=0$ holds. The point is that the derivation
$\,D^\th=x\!\.\!\pa+ 2\th\frac{\pa}{\pa\th}\,$ fulfills
$\,D^\th\th\ne0$, so it does not satisfy
\begin{equation*}
[D^\th x^1, x^2]_{\star_\Th} = [D^\th x^2 ,x^1]_{\star_\Th}\,,
\end{equation*}
which is equivalent to the integrability condition. Hence $D^\th$ is
outer. Analogous arguments work for our generators $G^\Th$.

\newpage
%\bibliography{apssamp}

%\end{document}

\end{document}